\documentclass[aps,prb,preprint,showpacs]{revtex4}

\usepackage{graphicx} 

\begin{document}

\title{Quantum Entanglement and Order Parameter in a Paired Finite
Fermi System}

\author{Z. Gedik}
\affiliation{Faculty of Engineering and Natural Sciences,
Sabanci University, 81474 Tuzla, Istanbul, Turkey}

\date{\today}

\begin{abstract}

We study the pairing correlations in a finite Fermi system from
quantum entanglement point of view. We investigate the relation
between the order parameter, which has been introduced recently to
describe both finite and infinite superconductors, and the
concurrence. For a proper definition of the concurrence, we argue
that a possible generalization of spin flip transformation is time
reversal operation. While for a system with indefinite number of
particles concurrence is a good measure of entanglement, for a
finite system it does not distinguish between normal and
superconducting states. We propose that the expectation value of
the radial operator for the total pseudospin can be used to
identify entanglement of pairing.

\end{abstract}


\maketitle

Entanglement is a fundamental quantum mechanical property
\cite{schrodinger} which plays a central role in the quantum
information theory \cite{bouwmeester}. On the other hand, proposed
measures of entanglement, including the entanglement of formation
which quantifies the resources needed to create a given entangled
state \cite{bennett}, are generally not very proper for analytical
calculations. Making use of a spin flip transformation, Wootters
\cite{wootters} introduced so-called the concurrence to simplify
the notion of entanglement of formation and Mart\'{i}n-Delgado
\cite{martindelgado} extended and applied his results to a
many-body problem, namely the BCS ground state of
superconductivity \cite{bardeen, bogolubov1}. In this work, we
examine the same problem for a finite system where the number of
fermions is fixed. For this purpose, we make use of an order
parameter proposed recently to describe both microcanonical and
grandcanonical superconductors \cite{gedik}.

Experimental works on superconducting metallic islands at
nanometer scale raised questions about pairing correlations
\cite{tuominen,lafarge,ralph}. For a bulk system,
superconductivity can be described by a complex order parameter
$\Delta$. The equations have the symmetry that if $\Delta$ is a
solution, then $e^{i\theta}\Delta$ is also a solution
\cite{bardeen, bogolubov1}. However, in a finite system with fixed
number of electrons, the order parameter $\Delta=\langle c_{-{\bf
k}\downarrow}c_{{\bf k}\uparrow}\rangle$ vanishes since the
operator does not conserve the number of fermions. Here, $c_{-{\bf
k}\downarrow}$ and $c_{{\bf k}\uparrow}$ are the annihilation
operators for time reversed states $\mid -{\bf
k}\downarrow\rangle$ and $\mid {\bf k}\uparrow\rangle$,
respectively. In this case, superconductivity can be identified by
nonvanishing number parity effect parameter $\Delta_P$ since the
ground state energy of the system increases or decreases,
depending upon whether the total number becomes odd or even, by
addition of a new electron \cite{matveev,flocard}. Recently, an
order parameter has been proposed to unify the order parameter
$\Delta$ of the bulk limit and the number parity effect parameter
$\Delta_P$ of the nanoscopic superconductors \cite{gedik}. Using
the pseudospin representation \cite{anderson,bogolubov2} and the
SU(2) phase states \cite{vourdas} a quantum phase has been defined
for a superconductor with discrete energy levels along with
modulus of the order parameter which becomes equal to $\Delta_P$.
As we go from the nanoscopic limit to the bulk superconductor it
has been shown that the number parity effect parameter and the
SU(2) phase go to the amplitude and the phase of the bulk order
parameter, respectively. On the other hand, we can think of the
long range order in the superconducting phase as an entangled
state of Cooper pairs. In this paper we first discuss how to
calculate concurrence as a measure of entanglement and we examine
the relation between entanglement and the order parameter. We show
that while for a system with indefinite number of fermions
concurrence is a good measure of entanglement, for a finite system
it does not identify the pairing correlations. As a possible
solution we propose the amplitude of our order parameter, which is
nothing but the expectation value of the radial operator of the
total pseudospin, to detect entanglement of Cooper pairs.

For a finite Fermi system, such as a nanoscopic superconductor,
energy levels are also finite and discrete and hence  we can use a
reduced form of the BCS model \cite{vondelft1} which was applied
in nuclear physics and which has an exact solution
\cite{richardson}. The model Hamiltonian is
\begin{equation}
H=\sum_{j,\sigma}\epsilon_j
c_{j\sigma}^{\dagger}c_{j\sigma}
-g\sum_{j,j'}
c_{j\uparrow}^{\dagger}c_{j\downarrow}^{\dagger}
c_{j'\downarrow}c_{j'\uparrow}
\label{ham}
\end{equation}
where $g$ is the pairing coupling constant for the time-reversed
states $\mid j\uparrow\rangle$ and $\mid j\downarrow\rangle$, both
having the energy $\epsilon_j$. Here, $c_{j\sigma}^{\dagger}$
($c_{j\sigma}$) is the creation (annihilation) operator for state
$\mid j\sigma\rangle$ where $j\in\{1,...,\Omega\}$ and
$\sigma\in\{\uparrow,\downarrow\}$. For the model Hamiltonian
introduced above it has been shown that there exists a number
parity effect, namely the ground state energy for even number of
electrons is lower in comparison to neighboring odd number states
\cite{mastellone,berger,braun,dukelsky} including degenerate case
\cite{kulik}.

The key point in Wootters' formulation of concurrence is the spin
flip transformation. For a pure state of a single qubit
$|\psi\rangle$ it is defined by
\begin{equation}
|\tilde{\psi}\rangle=\sigma_{y}|\psi^{*}\rangle
\end{equation}
where $|\psi^{*}\rangle$ is obtained from $|\psi\rangle$ by taking
complex conjugates of expansion coefficients and $\sigma_{y}$ is
Pauli spin matrix. For a single spin-$\frac{1}{2}$ particle, this
is nothing but the time reversal operation. The spin degree of
freedom that we discuss here shouldn't be mixed with the
pseudospin to be introduced below. For the many body case, a
natural extension of the spin flip operation is the time reversal
operation \cite{martindelgado}. The action of the time reversal
operator $U_T$ on the creation operator is
\begin{equation}
U_Tc_{j\sigma}^{\dagger}U_T^\dagger=c_{j-\sigma}^{\dagger}.
\label{trans}
\end{equation}
A similar relation holds for the annihilation operator. To find
the transformed state $|\tilde{\psi}\rangle$ (in the active
picture) we can simply rewrite the transformed Hamiltonian (in the
passive picture) and evaluate the corresponding eigenstate.

To define the order parameter, we introduce the pseudospin
variables \cite{anderson,bogolubov2}
\begin{equation}
s_j^z=\frac{1}{2}
\left(c_{j\uparrow}^{\dagger}c_{j\uparrow}
+c_{j\downarrow}^{\dagger}c_{j\downarrow}-1\right),\;\;\;\;\;
s_j^-=c_{j\downarrow}c_{j\uparrow}=\left(s_j^+\right)^\dagger
\end{equation}
which obey the fundamental commutation relations of the SU(2)
algebra
\begin{equation}
\left[s_i^+,s_j^-\right]=2\delta_{ij}s_j^z,\;\;\;
\left[s_i^z,s_j^\pm\right]=\pm\delta_{ij}s_j^\pm.
\end{equation}
It is possible to rewrite the model Hamiltonian as
\begin{equation}
H=\sum_{j}2\epsilon_j\left(s_j^z+\frac{1}{2}\right)
-g\sum_{ij}s_i^+s_j^-.
\label{sham}
\end{equation}
The mapping from the Fermi operators to the pseudospin operators
is possible as long as all single particle states are doubly
occupied. Since the original Hamiltonian (\ref{ham}) contains no
terms which couple a singly occupied level to others, the only
role of such states will be blocking from pairing interaction.
Therefore, the summations in Eqn.~(\ref{sham}) are over doubly
occupied and empty states only. Both the  above mapping and the
BCS wave function \cite{bardeen} lack proper antisymmetrization
due to separate treatment of singly occupied states, but since the
model Hamiltonian (\ref{ham}) does not involve any scatterings
into or out of such states, antisymmetrization with respect to
interlevel pair exchange and intrapair electron exchange is
sufficient.

Given SU(2) algebra, for example the one generated by the
components of the total pseudospin operator ${\bf s}=\sum_i{\bf
s}_i$, we can introduce \cite{vourdas} the radial operator defined
by
\begin{equation}
s_r=\sqrt{s^+s^-}
\end{equation}
and the exponential of the phase operator given by
\begin{equation}
E=\sum_{m=-s}^{m=s}\mid S;sm+1\rangle\langle S;sm\mid.
\end{equation}
Here, $\mid S;sm\rangle$ is simultaneous eigenstate of ${\bf s}^2$
and $s_z$ operators with eigenvalues $s(s+1)$ and $m$,
respectively. The label $S$ has been introduced to distinguish
them from the phase states to be defined below. For integer $s$ or
on the so called Bose sector, the eigenstate of $E$ with
eigenvalue $\exp(-i2\pi n/(2s+1))$ is evaluated to be
\begin{equation}
\mid \theta;sn\rangle=\frac{1}{\sqrt{2s+1}}\sum_{m=-s}^{m=s}
\exp\left[i\frac{2\pi n}{2s+1}m\right]\mid S;sm\rangle
\end{equation}
and a similar expression holds for half integer $s$ or in the Fermi sector.

In terms of the radial and the exponential of the phase operators
for the total pseudospin, it is possible to rewrite the
interaction part of the Hamiltonian (\ref{sham}) as
$-gs_rEE^\dagger s_r$. Since $E$ is unitary, we have
$EE^\dagger=I$ but we are going to keep $E$ and $E^\dagger$
without cancellation to introduce the phase properly. Now, we
define $\langle s_r \rangle$ and $\langle E\rangle$ as the
amplitude and phase of the order parameter, respectively. It has
been proven that $\langle s_r \rangle$ becomes identical to the
modulus of the BCS order parameter in the bulk limit while in the
nanoscopic limit it reduces to the number parity effect parameter
$\Delta_P$ \cite{gedik}. Furthermore, in the bulk limit, $\langle
E\rangle$ becomes identical to the phase of the BCS order
parameter. Next, we examine how the amplitude and the phase of the
order parameter is transformed under the time reversal operation
$T$. Equation (\ref{trans}) implies that the components of the
total pseudospin operator $\textbf{s}$ transform according to
\begin{eqnarray}
U_{T}s_{x}U_{T}^\dagger &=-s_{x}\\ \nonumber
U_{T}s_{y}U_{T}^\dagger &=-s_{y}\\
U_{T}s_{z}U_{T}^\dagger
&=s_{z}. \nonumber
\end{eqnarray}
The transformation has immediate consequences on the order
parameter. First, the amplitude $\langle s_r \rangle$ remains
unchanged. Second, since $\langle E\rangle=\langle
s^{+}\rangle/\langle \textbf{s}^{2}-s_{z}^{2}-s_{z}\rangle$, the
exponential of the phase expectation value acquires a minus sign
or phase angle change by $\pi$. This is consistent with the
prediction of BCS mean field treatment \cite{martindelgado}. Here,
we note that pseudospin operator does not transform exactly like
spin operator $\mathbf{\sigma}$ whose components satisfy
\begin{eqnarray}
U_{T}\sigma_{x}U_{T}^\dagger &=-\sigma_{x}\\ \nonumber
U_{T}\sigma_{y}U_{T}^\dagger &=\sigma_{y}\\
U_{T}\sigma_{z}U_{T}^\dagger &=-\sigma_{z} \nonumber
\end{eqnarray}
in the standard representation of Pauli matrices \cite{gottfried}.

For a given state $|\psi\rangle$, the central quantity concurrence
is defined by \cite{wootters}
\begin{equation}
C(\psi)=\mid\langle\psi\mid\tilde{\psi}\rangle\mid.
\end{equation}
Since $[s_r,s_z]=0$, $s_r$ leads to a good quantum number even for
a finite system. The eigenstates and including the ground state of
the model Hamiltonian will be of the form
\begin{equation}
\mid\psi_m\rangle=\sum_s c_s\mid S;sm\rangle \label{eig}
\end{equation}
because the interaction term commutes with ${\bf s}^2$ and $s_z$
while the single particle part commutes with the latter, only. In
general, the total spin in is multiply degenerate. We can
calculate the expectation value of the radial operator as
\begin{equation}
\langle s_r \rangle=\sum_s\mid c_s\mid^2\sqrt{s(s+1)-m(m-1).}
\end{equation}
Since the problem is exactly solvable, $c_s$ coefficients can be
found numerically \cite{vondelft2}. In terms of these coefficients
we can write down the transformed state which will be same as
(\ref{eig}) except that all coefficients will be replaced by their
complex conjugates. Then we evaluate the concurrence as
\begin{equation}
C(\psi)=\mid\sum_s c_s^2\mid.
\end{equation}
The BCS ground state, which is superposition of states of the form
(\ref{eig}) with different $m$ values, corresponds to phase states
$\mid\theta;sn\rangle$ in our notation \cite{gedik}. In other
words, it is an extended state in $m-$space and $C(BCS)$ can be
calculated explicitly \cite{martindelgado}. Here, we can evaluate
the same quantity for the phase states. However, since the phase
states are defined for a given $s$ value, we need to generalize
(\ref{eig}) by
\begin{equation}
\mid \psi_\theta\rangle\propto\sum_m e^{im\theta}\mid\psi_m\rangle
\end{equation}
from which we find that
\begin{equation}
C(\psi_\theta)\propto\mid\sum_s c_s^2\mid\mid\sum_m
e^{i2m\theta}\mid.
\end{equation}
The second term implies that concurrence vanishes in contrast to
the Fermi sea state. Hence, concurrence is a distinguishing
parameter for entanglement of Cooper pairs.

For a state with real expansion coefficients, assuming (\ref{eig})
is normalized, concurrence is unity, i.e. it is same as unpaired
state. This is the case for a system with fixed number of
fermions. A simple and analytically solvable example is a system
composed of a single, $d-$fold degenerate energy level
\cite{mottelson,bozat}. Therefore, microcanocial entanglement of
pairing, which we define by following Mart\'{i}n-Delgado as the
difference between concurrence values of the Fermi sea and the BCS
ground state, vanishes. Although concurrence does not distinguish
between the normal and the superconducting states, amplitude of
the order parameter $\langle s_r\rangle$ still identifies pairing
correlations and hence it can be used as a signature of
entanglement of Cooper pairs.

In conclusion, for a superconducting system with indefinite number
of particles, concurrence vanishes while it is unity for the Fermi
sea and hence it can be utilized to detect the existence of
entanglement. On the other hand, for a finite system it is unity
in the superconducting state, too. However, the order parameter of
superconductivity which we propose can still be used to identify
entanglement of Cooper pairs.

\begin{acknowledgments}
It is my pleasure to be a contributor in the Festschrift in the
honor of my ex-supervisor and colleague Prof. Salim Ciraci's 60th
birthday.

I acknowledge the hospitality of NIST Center for Neutron Research
at Gaithersburg where part of the work was performed. I also thank
to O. G\"{u}lseren and T. Y{\i}ld{\i}r{\i}m for helpful
discussions. This work was partially supported by the Turkish
Academy of Sciences T{\"U}BA/GEB{\.I}P Program.
\end{acknowledgments}

\end{document}